\documentclass[preprint,showpacs,superscriptaddress,nofootinbib]{revtex4}
\pdfoutput=1

\usepackage{amsfonts,amsmath,amssymb}
\usepackage{enumerate}
\usepackage{hyperref}
\usepackage{graphicx}
\usepackage{slashed}
\newcommand{\be}{\begin{equation}}
\newcommand{\ee}{\end{equation}}

\newcommand{\bea}{\begin{eqnarray}}
\newcommand{\eea}{\end{eqnarray}}

\newcommand{\bes}{\begin{subequations}}
\newcommand{\ees}{\end{subequations}}

\newcommand{\nn}{\nonumber}

\newcommand{\ra}{\rightarrow}

\newcommand{\D}{\Delta}
\newcommand{\tD}{{\tilde \D}}
\newcommand{\te}{\theta}

\begin{document}

\title{The Backreacted K\"ahler Geometry of Wrapped Branes}
\author{Nakwoo Kim}
\email{nkim@khu.ac.kr, nkim@ias.edu}
\affiliation{Department of Physics
and Research Institute of Basic Science, \\ Kyung Hee University, 
 Hoegi-dong, Dongdaemun-gu,
 Seoul 130-701, Korea}
\affiliation{School of Natural Sciences, Institute for Advanced Study, \\ Princeton, NJ 08540, USA}

\begin{abstract}
For supersymmetric solutions of D3(M2) branes with $AdS_3  (AdS_2)$ factor, it is known that the internal space is   expressible as $U(1)$ fibration over K\"ahler space which satisfies a specific partial differential equation involving the Ricci tensor.  In this paper we study the wrapped brane solutions of D3 and M2-branes which were originally constructed using gauged supergravity and uplifted to $D=10$ and $D=11$. We rewrite the solutions in canonical form, identify the backreacted K\"ahler geometry, and present a class of solutions which satisfy the Killing spinor equation.
\end{abstract}


\maketitle
\section{Introduction}
According to the AdS/CFT correspondence \cite{Maldacena:1997re}, given a supersymmetric solution of $D=10$ or $D=11$ supergravity with an anti-de-Sitter (AdS) factor we expect there should be a dual supersymmetric conformal field theory (SCFT). The maximally supersymmetric solutions arise from M5, D3, or M2 branes in the near horizon limit. A rich class of nontrivial solutions with less supersymmetries can be obtained when such branes are put at the tip of  a singular special-holonomy manifold. For 1/4-BPS and higher, the supergravity solutions are categorically written as a direct product of AdS and Sasaki-Einstein (SE) manifolds. The check of AdS/CFT duality involving a nontrivial SE manifold is a very actively studied research area these days. For IIB backgrounds $AdS_5\times SE_5$, essential references are \cite{Klebanov:1998hh}\cite{Martelli:2004wu}. For $AdS_4\times SE_7$ in M-theory, the identification of dual SCFT is not fully understood yet. The readers are referred to \cite{Aharony:2008ug} for M2-branes in ${\mathbb C}^4/\mathbb{Z}_k$ orbifold, while for nontrivial SE backgrounds see e.g. \cite{Closset:2012ep} for a recent study.

A more sophisticated method to obtain SCFT and dual AdS is to consider branes wrapped on a supersymmetric cycle in a special holonomy manifolds. 
Via topological twisting, the lower-dimensional field theory on the brane can retain supersymmetry. The twisting is best facilitated in the framework of lower-dimensional gauged supergravity \cite{Maldacena:2000mw}, and it is straightforward to uplift the explicit solutions to 10 or 11 dimensions. The dual field theories for wrapped branes have been mysterious for many years, especially for M5 and M2-branes. For M5-branes the $D=4$ SCFTs constructed in 
 \cite{Gaiotto:2009we} may provide the answer. 
It is illustrated there that M5-branes wrapped on a Riemann surface with punctures give rise to a variety of novel quiver gauge theories which flow to ${\cal N}=2$ SCFT in IR. It is argued in \cite{Gaiotto:2009gz} that the gravity dual in the large-$N$ limit should be ${\cal N}=2$ solutions of \cite{Maldacena:2000mw}, by comparing the central charges or the moduli space of marginal operators. 

On the other hand, one may systematically explore the consequence of unbroken supersymmetry and AdS geometry in supergravity solutions. Typically one starts with the existence of Killing spinor fields, and look for the form of supergravity solutions by analyzing the algebraic and differential identities between spinor bilinears. This method has been especially powerful with 1/2-BPS geometries \cite{Lin:2004nb}\cite{OColgain:2010ev}. For 1/4-BPS and below the form of the metric is rather involved, but with the use of ingenious ansatz one may sometimes find new solutions. In particular, inhomogeneous SE manifolds $Y^{p,q}$ were found this way \cite{Gauntlett:2004zh}\cite{Gauntlett:2004yd}. 

Some years ago the author applied this Killing spinor technique to the case of D3-branes geometry with an $AdS_3$ factor \cite{Kim:2005ez}. More concretely, we look for $AdS_3$ solutions in IIB supergravity where only the metric and five-form field strength are turned on. The result is quite elegant: the seven-dimensional internal space takes the form of twisted $U(1)$ fibration over six-dimensional K\"ahler space satisfying a higher-order differential equation involving Ricci tensor and scalar curvature.  
\be
\Box R - \frac{1}{2} R^2 + R_{ij} R^{ij} = 0 . 
\label{ke}
\ee
The warp factor for $AdS_3$ and the twisting of $U(1)$ can be fixed once the K\"ahler space is determined. 

Remarkably, one discovers basically the same structure for M2-brane configurations containing $AdS_2$ \cite{Kim:2006qu}. This time the internal space is $U(1)$ fibration over eight-dimensional space satisfying again \eqref{ke}. It was illustrated that
many known solutions with $AdS_3$ or $AdS_2$ factor can be rewritten this way, and also some new $AdS_3, AdS_2$ solutions were constructed by solving \eqref{ke} directly \cite{Gauntlett:2006ns,Kim:2007hv,Gauntlett:2007ts,Donos:2008ug}. (See also \cite{MacConamhna:2006nb,Figueras:2007cn,Gauntlett:2007ph,Colgain:2010wb} for general properties of wrapped D3 and M2-branes.)
It is also shown that this structure can be generalized to arbitrary higher dimensions  \cite{Gauntlett:2007ts}: $U(1)$ fibration over a solution of \eqref{ke} can be thought of a solution of a $(2n+1)$-dimensional bosonic action comprised of metric, one-form and a scalar field.  Existence of Killing spinor equation is reduced to \eqref{ke} for any $n\ge 3$.
It will be also worthwhile to comment that \eqref{ke} was recently reproduced as a special case of extremal black hole geometry in \cite{Gran:2011ak}.

In this paper we are interested in $AdS_3$ and $AdS_2$ solutions of Refs. \cite{Maldacena:2000mw} and \cite{Gauntlett:2001qs} which are respectively constructed from $D=5$ and $D=4$ gauged supergravity theories and interpreted as D3 and M2 branes wrapped on Riemann surface. We will first identify the six and eight-dimensional K\"ahler subspace which satisfy \eqref{ke}, from the wrapped brane solutions in \cite{Maldacena:2000mw,Gauntlett:2001qs} For the case of D3 wrapped on $H^2$ in Calabi-Yau 3-manifold (CY3), it turns out that the six-dimensional space is locally $H^2\times S^2$ fibred over a squashed $S^2$. Similarly, M2-branes wrapped on a $H^2$ 2-cycle in CY4 lead to a solution of \eqref{ke} which is locally $H^2\times KE_{2n}$ over a squashed $S^2$ ($KE^+_{2n}$ is K\"ahler-Einstein manifold with real dimension $2n$ and positive curvature).  Motivated by these solutions, we consider a general ansatz for \eqref{ke} and find solutions which are locally $H^2\times KE^+_{2n}$ over squashed $S^2$, which is our main result in this article. 

This paper is organized as follows. In section \ref{2} we review the Killing spinor geometry of  \cite{Gauntlett:2007ts} and explain how the solutions can be embedded in IIB and M-theory. 
In section \ref{3}, we consider the wrapped D3-brane solution of \cite{Maldacena:2000mw} and M2-brane solution in \cite{Gauntlett:2001qs} and identify the K\"ahler space satisfying \eqref{ke}.  In section \ref{4} we consider a higher dimensional generalization which takes a form of $H^2\times KE^+_{2n}$ over $S^2$. We conclude in section \ref{5}.

\section{BPS Geometry of $AdS_3$ from D3, and $AdS_2$ from M2}
\label{2}
Here we review the geometries studied in \cite{Kim:2005ez,Kim:2006qu,Gauntlett:2006ns} and generalized in \cite{Gauntlett:2007ts}. We are interested in the following Lagrangian density in $(2n+1)$-dimensions $(n\ge 3)$, 
\bea
{\cal L} &=& e^{(1-n)B} \left[
R + \frac{n(2n-3)}{2} (\nabla B)^2 + \frac{1}{4} e^{2B} F_{ab} F^{ab}
- \frac{2n}{(n-2)^2}
\right] . 
\label{la}
\eea
The field contents consist of the metric, a scalar $B$, and a vector field strength $F=dA$. We may consider the same action defined for a space with Lorentzian signature, but for our purpose here it is understood that the metric defines a Riemannian manifold.  

It is shown in \cite{Gauntlett:2007ts} that the above Lagrangian admits a particular type of Killing spinors. More concretely, we consider Killing spinor equations 
\bea
\left[ \Gamma^a \nabla_a B + i \frac{2(n-1)}{n-2} + \frac{1}{2} e^B F_{ab} \Gamma^{ab} 
\right] \eta &=& 0 , 
\nn\\
\left[ \nabla_c + \frac{i}{2} \Gamma_c + \frac{1}{8} e^B F_{ab} \Gamma_{c}^{\;\; ab} \right] \eta &=& 0 .  
\label{ks}
\eea
$\eta$ is a spinor field in $(2n+1)$-dimensions and $\Gamma_a$'s are generators of $\mbox{Cliff}(2n+1)$. As explained in \cite{Gauntlett:2006ns}, one can check that the existence of Killing spinor $\eta$, the Bianchi identity $dF=0$, and the gauge field equation $d( e^{(3-n)B} *F)=0$  guarantee that the equations of motion derived from \eqref{la} are satisfied. 

Furthermore, the above models describe in fact interesting subsectors of $D=10$ IIB supergravity and $D=11$ supergravity, for $n=3$ and $n=4$ respectively. Given a Killing spinor solution to \eqref{ks} for $n=3$, we can embed it in a $D=10$ IIB solution as follows: 
\bea
ds^2 &=& e^{-B/2} \left[ ds^2 ( AdS_3) + ds^2_7 \right] ,
\nn\\
F_5 &=& -\frac{1}{4} \left[ {\rm Vol}(AdS_3) \wedge F - *_7 F \right] . 
\label{10}
\eea
We have a similar mapping for $n=4$ case, with M-theory. 
\bea
ds^2 &=& e^{-2B/3} \left[ ds^2 ( AdS_2) + ds^2_9 \right] ,
\nn\\
G_4 &=& {\rm Vol}(AdS_2) \wedge F  . 
\label{11}
\eea
And of course the Killing spinor equations can be directly derived from the field ansatz and supersymmetry transformation rules of $D=10$ and $D=11$ supergravity. 

One can proceed further and study the consequence of Killing spinors on the form of metric and other fields. It is a standard Killing spinor technique which has been extensively studied in recent literature. It turns out that our $(2n+1)$-dimensional solution can be always written locally
\bea
ds^2_{2n+1} = c^2 (dz+P)^2 + e^B ds^2_{2n}
\label{odd}
\eea
i.e. as $U(1)$ vibration over $(2n)$-dimensional space. Here $c=n/2-1$, $\partial_z$ is a Killing vector and the $(2n+1)$-dimensional fields are independent of $z$. Most importantly it turns out that the $(2n)$-dimensional base space should be K\"ahler, and satisfy a {\it master equation} 
\be
\Box R - \frac{1}{2} R^2 + R_{ij} R^{ij} = 0 . 
\label{me}
\ee
Here $P$ is a one-form and related to K\"ahler geometry such that $dP$ is the Ricci two-form for $ds^2_{2n}$. The scalar field and the gauge field are then completely determined by the geometric data, as 
\bea
e^B &=& c^2 \left( \frac{R}{2} \right) ,
\nn\\
F &=& - \frac{1}{c} J + c \, d \left[ e^{-B} ( dz+ P ) \right] . 
\eea
where $J$ is the K\"ahler two-form of $ds^2_{2n}$.

As discussed in \cite{Gauntlett:2007ts} the physical application is not clear for $n>4$, but it is mathematically intriguing that there exists a class of {\it pseudo-supergravity} models for all $n\ge 3$ which includes physically interesting solutions for String/M-theory. 

\section{Wrapped brane solutions for D3 in CY3, and M2 in CY4}
\label{3}
\subsection{D3-branes wrapping a 2-cycle in CY3}
We will start with a specific $D=10$ IIB solution presented as D3-branes wrapped on a 2-cycle in Calabi-Yau 3-manifold, in \cite{Maldacena:2000mw}. The solution is best constructed in $D=7$ gauged supergravity first, and then uplifted to $D=10$. In $D=7$, the metric ansatz contains $H^2$ part and we also turn on magnetic flux on it. When acting on spinor fields, the spin connection of $H^2$ effectively cancel against the nonzero magnetic field, and this mechanism through AdS/CFT implies that that we have topological twist of boundary field theory. 

As given in Eq.(18) of \cite{Maldacena:2000mw}, the ten-dimensional metric is 
\bea
2^{3/2} ds^2 &=& \sqrt{\D} ds^2_{AdS_3} + \frac{1}{\sqrt{\D}}
\Bigg[ \D \, ds^2_{H^2} + 2 \D \,d\te^2 
\nn\\
&+& 4\cos^2\te\big( d\psi^2 + \sin^2\psi D\phi_1^2 +\cos^2\psi D\phi^2_2 \big)
+ 2\sin^2\te d\phi^2_3 \Bigg] , 
\label{mn}
\eea
where $\D=1+\sin^2\te$. In this subsection we denote $L^2=2^{-3/2}$. In our 
convention $H^2$ has unit radius, and $d(D\phi_1)=d(D\phi_2) = -\tfrac{1}{2} \mbox{Vol} (H^2)$. 

When we compare this metric with 
\eqref{10} and \eqref{odd}, it is obvious that $e^{-B} = L^4\Delta$ and the seven-dimensional internal space is to be rewritten as
\bea
&& 
\frac{1}{\D}
\Bigg[ \D \, ds^2_{H^2} + 2 \D \,d\te^2 
+ 4\cos^2\te\big( d\psi^2 + \sin^2\psi D\phi_1^2 +\cos^2\psi D\phi^2_2 \big)+ 2\sin^2\te d\phi^2_3 \Bigg] 
\nn\\
&&=
\frac{1}{4} ( dz+P)^2 + \frac{1}{L^4\Delta} ds^2_6 . 
\eea
 
   The constant-norm Killing vector $\partial_z$ must be a linear combination of $\partial_{\phi_i}, (i=1,2,3)$. We find it convenient to introduce a new set of coordinates as $ \phi=\phi_1-\phi_2, \varphi=\phi_1+\phi_2,\chi=2\psi$ and $z=2\phi_3$. The internal space metric and the twisting one-form $P$ are then found as  
\bea
\frac{1}{L^4} ds^2_6 &=& \Delta ds^2(H^2) + 2 \Delta d\theta^2 + \frac{\sin^2 2\theta}{2\Delta} (d\varphi - \cos\chi d\phi)^2 + \cos^2\theta ( d\chi^2 + \sin^2\chi d\phi^2) 
\label{6d}
\\
P&=& \frac{2}{\Delta} ( d\varphi - \cos \chi \phi ) . 
\eea
Now what we should check is that the $d=6$ metric in \eqref{6d} is K\"ahler and satisfies \eqref{me} and $dP$ is the Ricci two-form. This will be shown as a special case of a general construction in section \ref{4}.
 
\subsection{$AdS_2$ from wrapped M2 branes}
The prescription for supergravity solutions of wrapped branes \cite{Maldacena:2000mw} was generalized to M2-branes in \cite{Gauntlett:2001qs}.
It is convenient to consider a consistent subsector of $D=4, {\cal N}=8$ supergraivty with $U(1)^4$, and study $AdS_2\times H^2$ ansatz with magnetic flux on $H^2$. The solution we are interested in corresponds to M2-branes wrapped on a 2-cycle in CY 4-manifold presented in Eq.(3.19) of \cite{Gauntlett:2001qs}. Using the uplifting formula (see for instance Eq.(2.1) in \cite{Gauntlett:2001qs}), it is easy to check that 
the $D=11$ metric is 
\bea
ds^2_{11} &=& \frac{\tD^{2/3}}{2 \cdot 3^{5/3}\cdot e^2} \Biggr[ ds^2(AdS_2) + 3 ds^2(H_2) 
\nn\\
&& 
+ \frac{12}{\tD} \left(
3  \sum_{\alpha=1}^3 (d\mu^2_\alpha + \mu^2_\alpha D\phi_\alpha^2 )
+(d\mu^2_4 + \mu^2_4 d\phi_4^2) 
\right)
\Biggr] .
\eea
Here $\mu_a$ satisfy $\sum_{a=1}^4 \mu^2_a = 1$ and together with unconstrained angles $\phi_a \, (a=1,..,4)$ they parametrize $S^7$. The warp factor is $\tD = \mu^2_1 + \mu^2_2 + \mu^2_3  + 3 \mu^2_4$ and the coupling constant $e^2$ is related to the $D=4$ cosmological constant as $\Lambda_4 = - 12e^2$. The twisted one-forms satisfy $d(D\phi_\alpha) = -\frac{1}{3}{\rm Vol}(H^2)$ for $\alpha=1,2,3$.

To rewrite this metric in the form \eqref{11}, we first note that $e^{-B}=L^3 \tD$ with $L^{-2}= 2 \cdot 3^{5/3} \cdot e^2$. And in order to 
identify the constant norm Killing vector, we re-define some of angular coordinates as
\be
\phi_a \ra \phi_a + \phi_4/3, \quad a=1,2,3. 
\ee
Then one may easily check that the metric is written as \eqref{11} with 
\be
P = \frac{6}{\tD} \sum_{a=1}^3 \mu^2_a D\phi_a 
\ee
and the remaining $d=8$ K\"ahler base space is 
\bea
\frac{1}{L^3} ds^2_8 &=& 
 3\tD ds^2(H_2)
+ 12 (3 \sum_{a=1}^3 d\mu^2_a + d\mu^2_4 ) 
\nn\\
&& 
+ 36 \sum_{a=1}^3 \mu^2_a D\phi^2_a - \frac{36}{\tD}
\left( 
\sum_{a=1}^3 \mu^2_a D\phi_a \right)^2
\eea

It still looks complicated, but in fact one can rewrite this metric as $H^2\times \mathbb{CP}^2$ over squashed $S^2$, i.e. a natural generalization of \eqref{6d}. Let us define
\be
\mu_4 = \sin\theta, \quad \mu_a =  {\tilde\mu}_a \cos\theta \quad (a=1,2,3) . 
\ee 
And for $\phi_a$ angles we single out the diagonal part and define
\be
\phi_a = {\tilde\phi}_a + \psi \quad (a=1,2,3) . 
\ee
Note that $\sum_{a=1}^3 {\tilde\mu}_a^2 = 1 $ and $\sum_{a=1}^3 \tilde\phi_a = 0$. The Ricci potential is now 
\be
P = \frac{6\cos^2\theta}{\tD} \left( D\psi + \sum_{a=1}^3 {\tilde\mu}_a^2 d{\tilde\phi}_a
\right) , \quad d(D\psi) = - {\rm Vol} (H^2) . 
\ee
The metric becomes
\bea
\frac{1}{L^3} ds^2_8 &=& 
 3\tD ds^2(H_2)
+ 12 \tD d\theta^2 + \frac{27 \sin^2 2\theta}{\tD} (d\psi + \sum_{a=1}^3 {\tilde\mu}_a^2 d{\tilde\phi}_a)^2  
\nn\\
&& 
+36 \cos^2\theta \left( \sum_{a=1}^3 d{\tilde\mu}_a^2
+  \sum_{a=1}^3 \tilde\mu^2_a d\tilde\phi^2_a -
\left( 
\sum_{a=1}^3 \tilde\mu^2_a d\tilde\phi_a \right)^2
\right) . 
\label{8dfib}
\eea
We see that the part of the metric within big parentheses in the second line gives $\mathbb{CP}^2$. The easiest way to check this fact is to start with flat $\mathbb{C}^n$, parametrized by $z_j = \tilde\mu_j e^{i\tilde\phi_j}$ and rewrite the metric as metric cone of $S^{2n-1}$, which in turn is written as $U(1)$ fibration over $\mathbb{CP}^{n-1}$. 

The remaining task of checking \eqref{me} etc will be relegated to section \ref{4}, just as wrapped D3-brane case. 
\section{General Ansatz and Solutions}
\label{4}
Motivated by the explicit solutions in sectio n\ref{3}, we consider the following metric ansatz for \eqref{me}. 
\be
ds^2_{2n+4} = \D_1 ds^2(H_2) + \D_2 d\te^2 + \frac{\sin^2 2\te}{\D_2} D\psi^2 + \cos^2\te \,  ds^2(KE^+_{2n}) ,  
\label{ans1}
\ee
where $\Delta_1,\Delta_2$ are functions of $\theta$. 
For $n=1,2$ we expect to reproduce the wrapped D3 and M2-brane solutions explained in section \ref{3}. The nontrivial $U(1)$-fibration is defined as 
\be
D\psi = d\psi - P_{2} - P_{2n} . 
\ee
Here 
$P_2$ is Ricci potential for $H_2$, and $P_{2n}$ is the Ricci potential for $KE^+_{2n}$. We assume the overall normalization is such that 
\be
dP_{2} = -J_{2} , \quad dP_{2n} = + J_{2n}.
\ee
In other words, 
\be
 d(D\psi) = J_2-J_{2n}. 
\ee

To check the K\"ahler condition, we define the complex structure so that the K\"ahler 2-form is 
\be
J_{2n+4} = \D_1 J_2 + \sin 2\te d\te \wedge D\psi + \cos^2\te J_{2n} . 
\ee
One can easily check $J_{2n+4}$ is closed if $\D_1'=\sin 2\te$, i.e.
\be
\D_1 = c + \sin^2\te
\ee
where $c$ is constant. 

Now let us consider $(n,0)$-form $\Omega$. As it is well-known, we calculate $d\Omega$ and if it can be written as $d\Omega = i P \wedge \Omega$, the metric is K\"ahler with $dP$ the Ricci 2-form. We have
\be
\Omega = e^{-i\psi} \Omega_2 \wedge \cos^n\te \left(
\sqrt{\D_1\D_2} d\te + i \sqrt{\frac{\D_1}{\D_2}} \sin 2\te D\psi 
\right) \wedge \Omega_{2n} 
\ee
where we inserted the phase factor $e^{i\psi}$ for later convenience. This coefficient of course does not affect the result for Ricci-tensor. 
It is straightforward to check 
\bea
d\Omega_{2n+4} &=& i P_{2n+4} \wedge \Omega_{2n+4} ,
\\
P_{2n+4} &=& -\left[
1+ \frac{1}{\cos^n\te\sqrt{\D_1\D_2}}
\frac{d}{d\te}
\left(
\sqrt{\frac{\D_1}{\D_2}}\sin2\te \cos^n\te
\right)
\right] D\psi  \equiv f D\psi .
\eea
Summarizing, we have established that the ansatz \eqref{ans1} always gives us a locally K\"ahler space for any $\Delta_1=c+\sin^2\theta$ and arbitrary $\Delta_2(\theta)$. 

We can then compute the Ricci tensor by considering ${\cal R}_{2n+4} = dP_{2n+4}$. The computation is rather involved for general $\Delta_2$, since \eqref{me} is nonlinear and involves a fourth derivative. We have not been able to find the most general solution, but if we assume $\Delta_2$ is also a polynomial of $\sin^2\theta$, it is not difficult to check that the following is the only possibility for regular solutions: 
\be
\D_2 = 2n \D_1 = 2( n \sin^2\te +1) . 
\ee
For reference, we record that 
\bea
f = -\frac{2\cos^2\te}{1+n \sin^2\te} , \quad
R = \frac{4(1+n)}{1+n \sin^2 \te} . 
\eea
For $n=1$ this solution reduces to wrapped D3-branes, and for $n=2$ and $KE_4=\mathbb{CP}^2$ corresponds to M2 solutions presented in section \ref{3}. We can also take $KE_4 = S^2\times S^2$ and this gives us a physically distinct solution for a M2-brane configuration.  It will be also interesting to study the field theory dual of this new $AdS_2$ solution. 

\section{Discussion}
\label{5}
In this paper we have chosen the wrapped brane solutions constructed using gauged supergravity in \cite{Maldacena:2000mw}\cite{Gauntlett:2001qs}, and identified the back reacted K\"ahler geometry in terms of the general description in \cite{Gauntlett:2007ts}. The main motivation of this work was to obtain new AdS solutions using the generalized ansatz \eqref{ans1} which were drawn from wrapped brane solutions. By directly solving the master equation \eqref{me}, we were able to find the natural higher-dimensional generalization. 

The advantage of \eqref{me} is its generality, so in principle one can find any solution which probably cannot be obtained from lower-dimensional gauged supergravity system. In particular one can slightly change the ansatz of \eqref{ans1} and look for solutions. 
Consider the following metric ansatz
\be
ds^2_{2n+4} = \D_1 ds^2(KE_{2n}^-) + \D_2 d\te^2 + \frac{\sin^2 2\te}{\D_2} D\psi^2
+ \cos^2\te (d\chi^2+\sin^2\chi d\phi^2) 
\ee
where
\be
D\psi = d\psi + \cos\chi d\phi-P_{2n}, \quad dP_{2n} = - J_{2n} . 
\ee
For instance the space can be chosen as $ \otimes_{i=1}^n H^{2}_i\times S^2$ over squashed $S^2$. Now that we have increased the dimensionality of negatively curved part of the metric and we do not expect this class of metric may be covered by gauged supergravity. We have checked that this space allows K\"ahler structure provided $\D_1 = a + \sin^2\theta$, and computed Ricci tensor etc. Eq.\eqref{me} is again quite a complicated expression of $\D_2$ but one can proceed by assuming $\D_2$ is also a polynomial of $\sin^2\theta$. Unfortunately we have not found any physically interesting solution for $n>1$. We noticed for instance that $\D_1 = \sin^2\theta-\tfrac{3}{7}, \D_2=6\D_1$ locally satisfies \eqref{me} for $n=3$, but this space is singular since the hyperbolic space collapses at $\sin^2\theta=3/7$. 

Recently there is a revival of interest in the gravity solution of wrapped brane solutions. 
The authors of  \cite{Anderson:2011cz} considered a generalization of the gauged supergravity ansatz of \cite{Maldacena:2000mw} where the Riemann surface 2-cycle starts with an arbitrary metric in UV.  The motivation of \cite{Anderson:2011cz} was to prove a crucial assumption of \cite{Gaiotto:2009we} that for wrapped M5-branes the conformal factor of 2-cycle should be irrelevant in IR.  This can be  seen indeed from the elliptic flow equation derived in \cite{Anderson:2011cz}. But we would like to point out the central result of  \cite{Anderson:2011cz} was derived using $D=7$ gauged supergravity, not the original $D=11$ supergravity. It will be certainly desirable to study more general form of supersymmetric solutions for wrapped M5-branes (see e.g. \cite{Gauntlett:2006ux}) and study the flow equation, directly in $D=11$. 
On the other hand in  \cite{Bah:2011vv,Bah:2012dg}  new $AdS_5$ solutions and their field theory duals are constructed using general $\mathbb{C}^2$ bundle over Riemann surface.  If this construction can be also applied to D3 ro M2 branes it will be  interesting to identify the corresponding K\"ahler solution to \eqref{me}.  
\begin{acknowledgments}
This work is supported by the sabbatical leave petition program (2012) of Kyung Hee University, the National Research Foundation of Korea (NRF) funded by the Korean Government (MEST) with grant No. 2009-0085995, 2010-0023121, and also through the Center for Quantum Spacetime (CQUeST) of Sogang University with grant No. 2005-0049409.
\end{acknowledgments}

\bibliographystyle{JHEP}
\bibliography{flow}{}
\end{document}